\documentclass{elsart}
\usepackage{graphics}
\usepackage{epsfig}
\usepackage{amssymb}
\begin{document}
\newcommand{\EP}{\mbox{e$^+$}}
\newcommand{\EM}{\mbox{e$^-$}}
\newcommand{\EPEM}{\mbox{e$^+$e$^-$}}
\newcommand{\EMEM}{\mbox{e$^-$e$^-$}}
\newcommand{\EE}{\mbox{ee}}
\newcommand{\GG}{\mbox{$\gamma\gamma$}}
\newcommand{\GP}{\mbox{$\gamma$e$^+$}}
\newcommand{\GE}{\mbox{$\gamma$e}}
\newcommand{\LGE}{\mbox{$L_{\GE}$}}
\newcommand{\LGG}{\mbox{$L_{\GG}$}}
\newcommand{\LEE}{\mbox{$L_{\EE}$}}
\newcommand{\XG}{\mbox{$x_{\gamma}$}}
\newcommand{\TEV}{\mbox{TeV}}
\newcommand{\WGG}{\mbox{$W_{\gamma\gamma}$}}
\newcommand{\GEV}{\mbox{GeV}}
\newcommand{\EV}{\mbox{eV}}
\newcommand{\CM}{\mbox{cm}}
\newcommand{\M}{\mbox{m}}
\newcommand{\MM}{\mbox{mm}}
\newcommand{\NM}{\mbox{nm}}
\newcommand{\MKM}{\mbox{$\mu$m}}
\newcommand{\E}{\mbox{$\epsilon$}}
\newcommand{\EN}{\mbox{$\epsilon_n$}}
\newcommand{\EI}{\mbox{$\epsilon_i$}}
\newcommand{\ENI}{\mbox{$\epsilon_{ni}$}}
\newcommand{\ENX}{\mbox{$\epsilon_{nx}$}}
\newcommand{\ENY}{\mbox{$\epsilon_{ny}$}}
\newcommand{\EX}{\mbox{$\epsilon_x$}}
\newcommand{\EY}{\mbox{$\epsilon_y$}}
\newcommand{\SEC}{\mbox{s}}
\newcommand{\CMS}{\mbox{cm$^{-2}$s$^{-1}$}}
\newcommand{\MRAD}{\mbox{mrad}}
\newcommand{\IND}{\hspace*{\parindent}}
\newcommand{\beq}{\begin{equation}}
\newcommand{\eeq}{\end{equation}}
\newcommand{\beqn}{\begin{eqnarray}}
\newcommand{\eeqn}{\end{eqnarray}}
\newcommand{\dst}{\displaystyle}
\newcommand{\bm}{\boldmath}
\newcommand{\BX}{\mbox{$\beta_x$}}
\newcommand{\BY}{\mbox{$\beta_y$}}
\newcommand{\BI}{\mbox{$\beta_i$}}
\newcommand{\SX}{\mbox{$\sigma_x$}}
\newcommand{\SY}{\mbox{$\sigma_y$}}
\newcommand{\SZ}{\mbox{$\sigma_z$}}
\newcommand{\SI}{\mbox{$\sigma_i$}}
\newcommand{\SIP}{\mbox{$\sigma_i^{\prime}$}}
\newcommand{\n}{\mbox{$n_f$}}
\begin{frontmatter}
\date{} \title{Problems of multi-TeV photon colliders\thanksref{talk}}
\author{Valery Telnov \thanksref{cor}} 
\thanks[talk]{Talk at International Workshop on High-Energy Photon Colliders,
 Hamburg, Germany, 14-17 Jun 2000.  Submitted to Nucl.Instrum.Meth.A}
\thanks[cor]{e-mail: telnov@inp.nsk.su, telnov@mail.desy.de}
\address{Institute of Nuclear Physics, 630090 Novosibirsk, Russia \\
and DESY, Notkestr.85, D-22603 Hamburg, Germany}

\begin{abstract}
 
  A high energy photon collider (\GG, \GE) based on backward Compton
  scattering of laser light is a very natural supplement to \EPEM\ a
  linear collider and can significantly enrich the physics program.  The
  region below about one 0.5--1 TeV is very convenient from a technical
  point of view: wave length of the laser should be about 1 $\mu$m, i.e. in
  the region of most powerful solid state lasers, collision effects
  do not restrict the \GG\ luminosity.  In the multi-TeV energy region
  the situation is more complicated: the optimum laser wave length
  increases in proportionally with the energy, the required flash energy also
  increases due to nonlinear effects in the Compton scattering; bunch
  trains are shorter (for warm high gradient linacs), this leads to
  higher backgrounds; the collision effects (coherent pair \EPEM\ pair
  creation) restrict the luminosity. These problems and possible
  solutions are discussed in this paper.  A method of laser focusing
  is considered which allows the decrease of the required laser flash energy
  and the practical elimination of the problem of nonlinear effects in
  Compton scattering; a way to reduce collision effects and obtain ultimate 
  \GG\  luminosities at multi-TeV photon colliders is suggested.

\vspace*{0cm}
\end{abstract}
\end{frontmatter}

\section{Introduction}

The next generation of linear colliders JLC~\cite{JLC},
NLC~\cite{NLC}, TESLA~\cite{TESLA} are being developed at the energies
from 100 GeV to about 1 TeV. In this energy region, \EPEM\ linear
colliders are the best machines for study of elementary particles;
they can have a sufficient luminosity, good monochromaticity, and
rather low backgrounds.  Though already at these energies attainable
\EPEM\ luminosity is determined by collision effects.  In order to
reduce beamstrahlung during the beam collisions very flat beams should
be used with a vertical size of several nm. To obtain a sufficient
luminosity one has to increase average beam current, as a result of
which the total power consumption approaches 100 MW.

At multi-TeV linear colliders, such as CLIC~\cite{CLIC,Burkhardt}, all
problems are much more severe.  The required luminosity should vary
proportionally to $E_0^2$ (as soon as cross sections $\propto 1/E_0^2$).
To get a sufficient luminosity at a reasonable beam power one has to
further decrease beam sizes and admit larger energy spread due to
beamstrahlung, and increase total power consumption.  So, the problem of
multi-TeV \EPEM\ colliders is not just the acceleration of beams up to
a high energy, besides there are many other problems caused by collisions
effects.  As a result, the maximum energy of linear colliders (with
a reasonable luminosity) is about 2E $\sim$ 5 TeV,  namely such as
is considered in the CLIC project.

Photon colliders~\cite{GKST81,GKST83,GKST84} are based on the Compton
scattering of laser light on high energy electrons. This is only possible
at linear colliders where the beams are used only once.  Photons
are neutral particles therefore there is no beamstrahlung nor beam
instabilities. So, at first sight, it seems that in \GG\ colliders can
be optimized completely differently and perhaps, in \GG\ mode of
operation, linear colliders can reach higher energies.

In this paper I will try to give answers  these questions and also consider
the  problem of lasers  for multi-TeV photon colliders.
Consideration of the general principles of photon colliders and the present
status can be found
elsewhere~\cite{GKST83,GKST84,TEL90,TEL95,Brinkmann,TELTESLA}.  Though
the present paper focuses on problems of multi-TeV colliders the
results are valid and useful for all energies.

\section{Conversion of electrons to high energy photons}

At multi-TeV energies there are several complications connected with
the conversion of electrons to high energy photons:
\begin{enumerate}
\item  the optimum laser wave length  increases proportionally to the
electron energy; 
\item  nonlinear effects in the Compton scattering become more important; 
\item due to nonlinear effects the required flash energy is higher; 
\item it is not so clear as to what kind of laser can be used.  
\end{enumerate}
\subsection{Laser wave length}

\begin{figure}[!hbt]
\centering
\vspace*{0.cm}
\epsfig{file=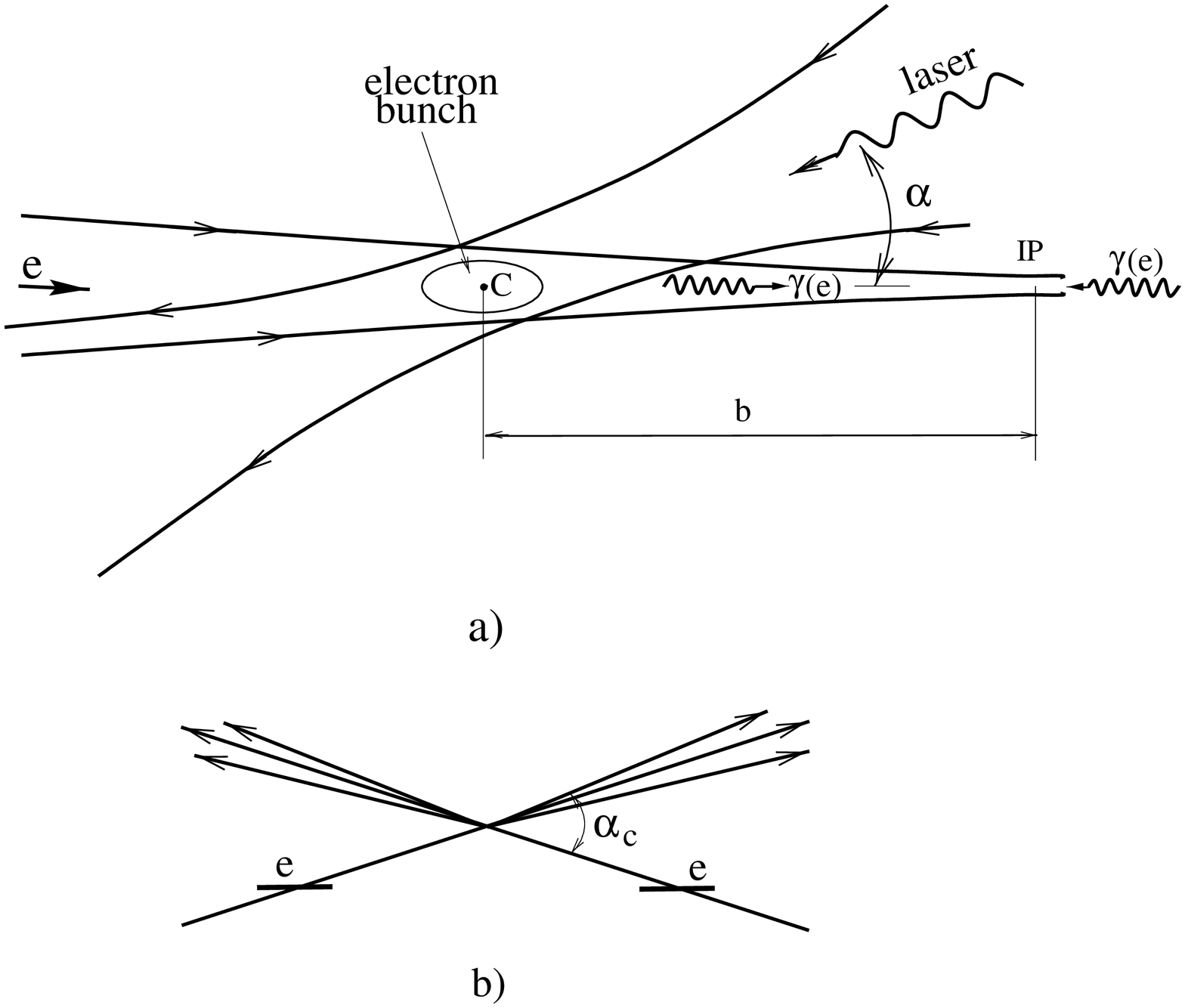,width=9cm,angle=0}
\vspace*{-0.cm}
\caption{Scheme of  \GG, \GE\ collider.}
\label{ggcol}
\end{figure}

 The maximum energy of the scattered photons is \cite{GKST83}
\begin{equation}
\omega_m=\frac{x}{x+1}E_0; \;\;\;\;
x = \frac{4E_0\omega_0\cos^2{\alpha/2}}{m^2c^4}
 \simeq 15.3\left[\frac{E_0}{\TEV}\right]
\left[\frac{\omega_0}{eV}\right],
\label{omegam}
\end{equation}
where $E_0$ is the electron energy, $\omega_0$ the energy of the laser
photon, see Fig.\ref{ggcol}a.  

\begin{figure}[!hbt]
\centering
\vspace*{-1.1cm}
\epsfig{file=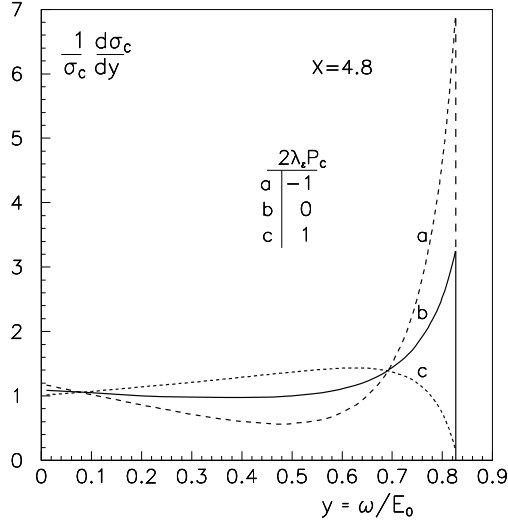,width=9cm,angle=0}
\vspace*{-1.cm}
\caption{Spectrum of  Compton scattered photons for various 
polarization of laser and electron beams~\cite{TEL95}.}
\label{spectra}
\end{figure}

The Compton cross section depends on
the longitudinal electron polarization $\lambda_e$ ($|\lambda_e| \leq
1/2)$ and  the circular polarization of laser photons $P_c$ 
\begin{equation}
\sigma_c = \sigma_{c,np} +2\lambda_e P_c \sigma_{c,1}.
\label{sigmac}
\end{equation}
Expressions for these cross sections can be found
elsewhere~\cite{GKST84,TEL95,Serbo}. The energy spectrum of the scattered
photons also depends on the product $2\lambda_e P_c$. A typical
spectrum of scattered photons for $x=4.8$ is shown in
Fig.\ref{spectra}.  One can see that in the case
$2\lambda_e P_c = -1$ (curve a) the energy spectrum has a high energy
peak at the maximum energy, while in the case $2\lambda_e P_c = 1$
(curve b) the distribution is flat and even approaches to zero at the maximum
energies.

With increasing $x$ the spectrum corresponding to $2\lambda_e P_c =
-1$ becomes  narrower, while the spectrum for $2\lambda_e P_c = 1$
remains quite wide, though the fraction of photons in the 
high energy part increases: at $x=4.8$ about half of the
photons have energy above $0.5\omega_m$, while at $x=50$  half of the photons
have energy above $0.8\omega_m$. Corresponding graphs
for large $x$ can be found in Ref.~\cite{Galynskii} in these
proceedings.

\begin{figure}[!hbt]
\centering
\vspace*{-1.cm}
\epsfig{file=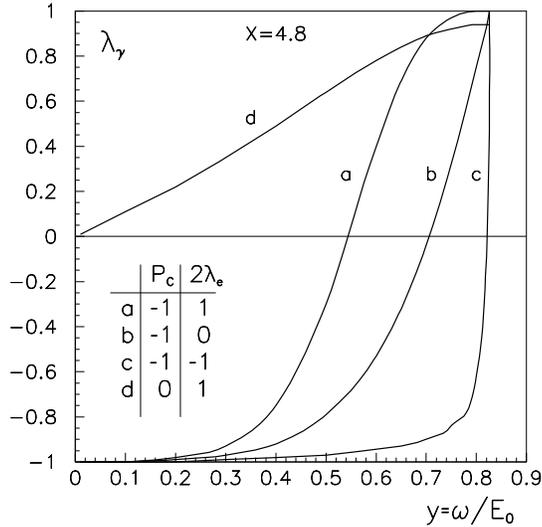,width=9cm,angle=0}
\vspace*{-1.cm}
\caption{Helicity of the Compton scattered photons vs $\omega/E_0$ for various 
polarizations of laser and electron beams~\cite{TEL95}.}
\label{polarization}
\end{figure}

Typical curves for polarization of the scattered photons  for $x=4.8$
are shown in Fig.\ref{polarization}.  For further discussion the
following features are important: in both cases a) and c)
(same as in the previous figure) the polarization of the  highest
energy photons  ($\omega = \omega_m$) is 100\% and opposite to the
polarization of the laser photons. 
However, for the case a) ($2\lambda_e
P_c = -1$) most of high energy photons (see Fig.~\ref{spectra}) have the
polarization $\lambda_{\gamma} \approx - P_c$, while for the case c)
($2\lambda_e P_c = 1$) only very small number of photons with
$\omega \approx \omega_m$ have $\lambda_{\gamma} \approx - P_c$, while most
of photons have the opposite polarization $\lambda_{\gamma} \approx P_c$.

\begin{figure}[!hbt]
\centering
\vspace*{-1.1cm}
\epsfig{file=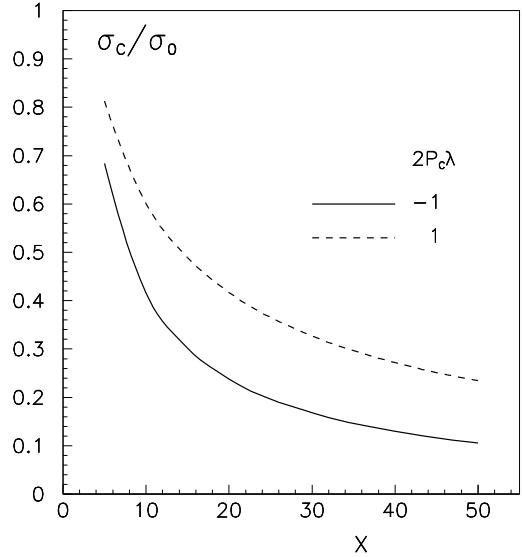,width=8.5cm,angle=0}
\vspace*{-0.5cm}
\caption{Compton cross section vs $x$ for two combinations of laser 
and electron polarizations.}
\label{crosssection}
\end{figure}

Dependence of the Compton cross section on $x$ for two cases of
polarization is shown in Fig.\ref{crosssection} in units of
$\sigma_0=\pi r_e^2$, where $r_e=e^2/mc^2=2.8 \times 10^{-13}$ cm
is the classical radius of the electron. Note, in this section we only
consider  the case of the {\it linear} Compton effect, when only
one laser photon is scattered from an electron. In the next section, we
note that for multi-TeV photon colliders nonlinear effects
(simultaneous interaction of the electron with several laser photons)
are very important.

With an increase in $x$ the energy of the back-scattered photons 
grows and the energy spectrum becomes  narrower, however, at $x > 4.8$ the
high energy photons may be lost due to creation of \EPEM\ pairs in the
collisions with laser photons~\cite{GKST83,TEL90,TEL95}.

The cross section for \EPEM\ production in a two-photon collision is
given by~\cite{Ispirian,Budnev}
\begin{equation}
\sigma_{\GG \to \EPEM} = \sigma_{np} + \lambda_1\lambda_2 \sigma_1,
\label{sigmae+e-}
\end{equation}
$$
 \sigma_{np} = \frac{4\sigma_0}{\XG}\left[2 \left(1 + \frac{4}{\XG} -
\frac{8}{\XG{^2}}\right)\ln{\frac{\sqrt{\XG} + \sqrt{\XG\ -4}}{2}} -
\left(1+ \frac{4}{\XG}\right)\sqrt{1-\frac{4}{\XG}}\;\right],
$$
\begin{equation}
 \sigma_1 = \frac{4\sigma_0}{\XG}\left[2\ln{\frac{\sqrt{\XG} + 
\sqrt{\XG\ -4}}{2}} - 3\sqrt{1-\frac{4}{\XG}}\;\right]
\label{sigmae+e-1},
\end{equation}
where $\XG\ = 4\omega_m \omega_o/ m^2c^4$ or $\XG\ = x^2/(x+1)$, and
$\lambda_1, \lambda_2$ are photon helicities.

\begin{figure}[!hbt]
\centering
\vspace*{-0.8cm}
\epsfig{file=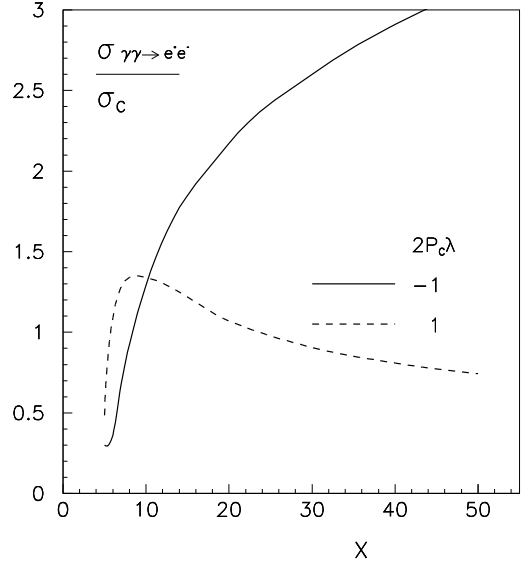,width=8.5cm,angle=0}
\vspace*{-0.4cm}
\caption{Ratio of cross sections for \EPEM\ pair creation in the collision
  of laser and high energy photon  and for the  Compton scattering.}
\label{ratio}
\end{figure}

The ratio $\sigma_{\GG \to \EPEM}/\sigma_c$ is shown in
Fig.\ref{ratio}.  Calculating this ratio we took
$\lambda_1\lambda_2=-1$ for $2\lambda_e P_c = -1$ and
$\lambda_1\lambda_2=1$ for $2\lambda_e P_c = 1$ (this was explained earlier).

\begin{figure}[!hbt]
\centering
\vspace*{-0.9cm}
\epsfig{file=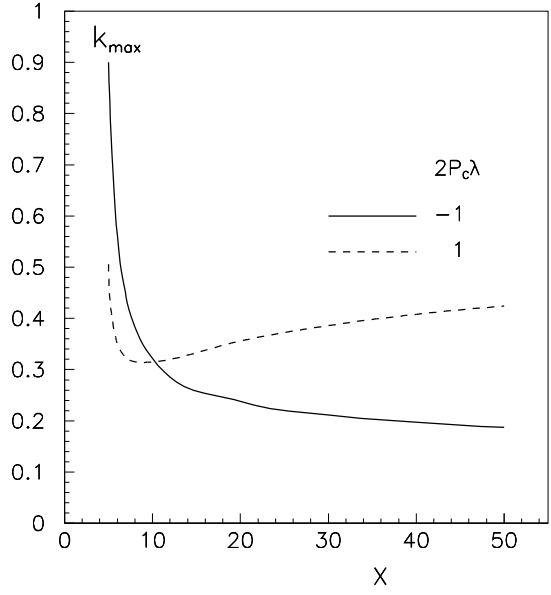,width=9cm,angle=0}
\vspace*{-0.4cm}
\caption{Maximum conversion coefficient vs $x$ for two combinations of laser
and electron polarizations, see comments in text. }
\label{kmax}
\end{figure}

Creation of high energy photons in the conversion region and its
conversion to \EPEM\ pairs is described by the corresponding kinetic
equation~\cite{TEL90}. The maximum yield of the high energy
photons from the conversion region (under the assumption that all photons
have energy close to $\omega_m$) is equal to
\begin{equation}
k_{max} = \frac{N_{\gamma\;max}}{N_0} = 
\frac{1}{p-1}(p^{1/(1-p)} -p^{p/(1-p)}),
\end{equation} 
where $p = \sigma_{\GG \to \EPEM}/\sigma_c$. Dependence of the maximum
conversion coefficient on $x$ is shown in Fig.\ref{kmax}. For $x<4.8$,
$k_{max}=1$ (in principle), though it would be more reasonable to assume the
thickness of the laser target to be equal to one collision length,
this gives $k_{max}=1-\exp{(-1)}=0.632$. For $x=50$ ($E_0 = 2.5$ TeV,
$\lambda = 1$ \MKM) $k_{max}$ is equal to 0.425 (0.187) for
$2\lambda_e P_c = 1$ $(-1)$, respectively. The \GG\ luminosity is
proportional to $k^2$, therefore, at $x=50$ it will be lower than at
$x < 4.8$ by a factor of 2.2 (11.4), for the two cases, respectively.
We see that the reduction of the luminosity for the case $2\lambda_e
P_c = 1$ is not dramatic and for multi-TeV photon colliders one can use 
lasers with $\lambda \sim 1\;\MKM$, which are optimal for $2E=500$ GeV 
($x \sim 4.8$).

  So, there are two possibilities for multi-TeV photon colliders :

1) $x \sim 4.8$ ($\lambda \sim 4E_0[\TEV]$ \MKM),  $2\lambda_e P_c = -1$.
 
2) $\lambda \sim 1$ \MKM, $2\lambda_e P_c = 1$, $\LGG\ \sim
0.4\LGG(x=4.8)$

Note, that all this is valid only in the case of  linear Compton
scattering.  Below we will see that, in the second case, the picture is
more complicated due to nonlinear effects in a strong field of the
laser wave.

\subsection{Nonlinear effects in Compton scattering}

\begin{figure}[!hbt]
\centering
\vspace*{-1.cm}
\epsfig{file=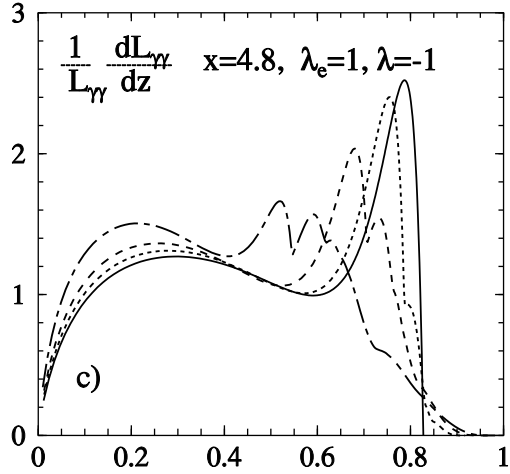,width=9cm,angle=0}
\vspace*{-1.5cm}
\caption{ Luminosity distributions for various values of $\xi^2$ at $x=4.8$ and
  $2\lambda_e P_c = -1$. Counting from the right the curves correspond
  to $\xi=0, 0.3, 1, 3$.}
\label{lm48_2ta}
\end{figure}

The field in the laser wave at the conversion region is very strong, so
that the electron (or high energy photon) can interact simultaneously
with several laser photons. These nonlinear effects are characterized
by the parameter~\cite{Berestetskii}

\begin{equation}
\xi^2 = \frac{e^2\overline{B^2}\hbar^2}{m^2c^2\omega_0^2} = 
\frac{2 n_{\gamma} r_e^2 \lambda}{\alpha},
\label{xi^2}
\end{equation}
where $n_{\gamma}$ is the density of laser photons. At $\xi^2 \ll 1$
the electron is scattered on one laser photon, while at $\xi^2 \gg 1$
-- on several.  Nonlinear effects in Compton scattering at photon
colliders are considered in detail in Ref.\cite{Galynskii} in these
proceedings.  Shortly, with grows of $\xi^2$ the Compton spectrum
becomes wider and is shifted to lower energies. Evolution of \GG\ 
luminosity distributions with the increase of $\xi^2$ at $x=4.8$ and
$2\lambda_e P_c = -1$ is shown in Fig.\ref{lm48_2ta} from
Ref.~\cite{Galynskii}.  One can see the shape of the luminosity
spectra are still acceptable up to $\xi^2\sim 0.5-1$.  For $x = 50$ and
$2\lambda_e P_c = 1$ (see Figs in Ref.\cite{Galynskii} the spectrum
does not change too much up to $\xi^2 < 3$.
Sensitivity of the Compton spectrum to the nonlinear effects can be
estimated as following. Due to transverse motion in the
electromagnetic wave the electron has the effective mass $m^2 =
m_e^2(1+\xi^2)$~\cite{Berestetskii}. As a result the maximum energy of
the scattered photons is decreased
\begin{equation}
\omega_m = xE_0/(x+1+\xi^2).
\end{equation} 
The relative shift of $\omega_m$  
\begin{equation}
\Delta \omega_m/\omega_m \sim \xi^2/(x +1 )
\end{equation} 
 is smaller for larger $x$.
 
 Beside worsening of the luminosity spectrum, large $\xi^2$ also leads 
 to decrease of the polarization in the maximum of the luminosity
 spectrum.  Partially, this is connected with the fact that photons of
 the same energy can be produced in the region with high or low
 $\xi^2$. In the later case, the polarization will be lower, because the
 ratio $\omega/\omega_m$ is lower while the polarization is high only
 for $\omega$ close to $\omega_m$.

One of the main practical consequences of the nonlinear effects is 
the increase of the required laser flash energy. This problem is considered in
the next section.

\subsection{Laser flash energy}

The collision probability of an electron with  laser photons in the
scheme shown in Fig.\ref{ggcol} can be estimated as follows
\begin{equation}
p \approx \frac{2\sigma_c n_0}{\theta} 
\int_{-\infty}^{\infty} e^{-\frac{x^2}{2\sigma_x^2}}dx = 
\frac{2\sqrt{2\pi}\sigma_c n_0 \sigma_x}{\theta},
\label{p1}
\end{equation} 
where $n_0$ is the density of laser photons at the focal point,
$\sigma_x$ is the r.m.s. transverse size of the laser beam in the
conversion region (it is assumed to be constant), $\theta$ is the
collisions angle.  The r.m.s. beam size can be expressed via the beam
divergence
\begin{equation}
\sigma_x = \frac{\lambda}{4\pi\sigma_{x^{\prime}}},
\label{sigmax}
\end{equation} 
(this follows from the Heisenberg uncertainty principle:
$\sigma_{p_x}\sigma_x= \hbar/2$).  To confine a Gaussian laser
beam,~\footnote{From the practical point of view the flat shape may be
   better, but here we consider  Gaussian beams, for simplicity } the
 collision angle $\theta$ should be several times larger than the beam
divergence
\begin{equation}
\theta = s \sigma_{x^{\prime}} = \frac{s \lambda }{4\pi\sigma_x},
\label{theta}
\end{equation}  
where $s \sim 2$ (mirrors cover $1-e^{-2}=0.865$ of the Gaussian
beam).\footnote{this is not the loss of the 13\% of energy, the laser
  beam can have any profile (superposition of many transverse modes),
  this number just means that formulas for a Gaussian beam describe
  the density distribution near the laser focus with a sufficient
  accuracy.}  Substituting $\sigma_x$ from (\ref{theta}) and $n_0$
from (\ref{xi^2}) into (\ref{p1}) we get the collision probability
\begin{equation}
p \approx \frac{\sqrt{2\pi}\alpha\sigma_c\xi^2 s}{4\pi\theta^2 r_e^2}.
\label{p2}
\end{equation}
This is a useful relation between the conversion probability, angular size of
the focusing system ($\pm \theta$) and parameter $\xi^2$.
Substituting (\ref{theta}) to (\ref{p2}) we get the radius of the laser beam
\begin{equation}
\sigma_x^2 \approx 
\frac{r_e^2\lambda^2 s p}{4\pi\sqrt{2\pi}\alpha\sigma_c\xi^2}
\label{sigmax2}.
\end{equation}

The distribution of photons in a Gaussian beam with uniform
longitudinal density is given by 
\begin{equation}
dN = \frac{N}{2\pi l_{\gamma}\sigma_x\sigma_y}e^{-x^2/2\sigma_x^2
-y^2/2\sigma_y^2}dx dy dz,
\label{dN}
\end{equation}   
where $l_{\gamma}$ is the bunch length.  Hence, for a round beam
$N=2n_0\pi \sigma_x^2 l_{\gamma}$, where $n_0$ is the photon density
at the focal point, which can be expressed via the parameter $\xi^2$
using Eq.(\ref{xi^2}).  The laser flash energy
\begin{equation}
A= N\omega_0 = 2\pi\sigma_x^2 n_0 \omega_0 l_{\gamma},
\label{A1}
\end{equation}
where the length of the laser bunch follows from the requirements that
laser photons should be present in the conversion region all the time
until an electron bunch crosses the region $\pm \sigma_x/\theta$, this
gives
\begin{equation}
l_{\gamma} = 4\sigma_x/\theta + l_e = 16\pi\sigma_x^2/\lambda s + l_e.
\label{le}
\end{equation}
Substituting (\ref{xi^2},\ref{sigmax2},\ref{le}) to (\ref{A1}) we get
the required laser energy
\begin{equation}
A= \frac{mc^2l_eps}{\alpha r_e \sqrt{8\pi}}\left(\frac{\sigma_0}
{\sigma_c}\right)\left[1+ \frac{4\lambda p}{\sqrt{2\pi}\pi\alpha l_e 
\xi^2}\left(\frac{\sigma_0}{\sigma_c}\right)   \right].
\label{A2}
\end{equation}
For $s=2$ the flash energy is equal to
\begin{equation}
A_0 [J]= 16p\left(\frac{\sigma_0}
{\sigma_c}\right)l_e\left[1+p\frac{70\lambda \sigma_0}{\xi^2 l_e 
\sigma_c}\right],
\label{A3}
\end{equation}
where all lengths are expressed in cm. Comparison with the simulation shows
that this formula works better with somewhat larger coefficients: 16$\to$20
and 70 $\to$ 150. We will use these numbers for further estimations.
 The first term in
Eqs.\ref{A2},\ref{A3} is some minimum flash energy determined by the
diffraction of laser beams, the second term is due to the nonlinear
effects (when the maximum value of the parameter $\xi^2$ is fixed).
With the increase of the electron beam energy the second term becomes
more important for three reasons: 1) if $x=$const, then $\lambda
\propto E_0$, 2) if $\lambda =$ const, then $x \propto E_0$ and the
ratio $\sigma_c/\sigma_c$ decreases (this increases the first term as
well), 3) the electron bunch length is smaller for accelerators with
higher acceleration gradient.  So, to have a reasonable laser flash
energy at multi-TeV photon colliders one has to increase the value of
the parameter $\xi^2$, this leads to a worsening of the luminosity
spectrum.\footnote{Section~\ref{new} shows how 
 this problem can be overcome.}

The required flash energy and contribution of each
term for three cases are given in table \ref{tflash}.
\begin{table}[!hbtp]
\caption{Laser flash energies for three sets of  parameters.}
\vspace{5mm}
{\renewcommand{\arraystretch}{1.2}
\begin{center}
\begin{tabular}{l c c c} \hline 
 & a & b & c  \\ \hline
$2E_0$, GeV & 500 & 5000 & 5000 \\ 
$2P_c\lambda_e$ & -1  & -1  & 1    \\
$\lambda$,\MKM\ & 1  & 10  & 1    \\
$\xi^2$ & 0.3  & 1  & 3    \\
$\sigma_c/\sigma_0$ & 0.7$\cdot$0.95  & 0.7$\cdot$0.89   
& 0.235$\cdot$0.68  \\  
$l_e,\MKM\ $ & 600   & 200   & 200     \\  
$p$ & 1  & 1  & 0.75 (close to opt.) \\ \hline
$A$, J & 1.8(1+1.25)=4  &0.64(1+12)=8.3   &1.9(1+1.17)=4.1  \\   
\end{tabular}
\end{center}
}
\label{tflash}
\end{table}   
Here, in the line $\sigma_c/\sigma_0$ the second number after
``$\cdot$'' is due to nonlinear effects. Let me remind also that in
the case c) the luminosity is lower by a factor of
2.3 than in the case b)  (for small $\xi^2$) due to \EPEM\ production.

So, in both cases of multi-TeV colliders b) and c), the flash energy
seems acceptable. Of course, in the case b) it is better to have
$\xi^2<0.3$ but in this case the flash energy would be too large, about 25
J.  

One more remark concerning the case c). Here we have assumed $\xi^2
=3$, but, in the case of large $x$ and large $\xi^2$, the linear
theory of \EPEM\ production in collisions between laser and high
energy photons may be not be valid due to the coherent \EPEM\ 
production~\cite{CHTEL,TEL90}.  This problem needs accurate study,
here I would just like to give some comment. The coherent pair
creation in the uniform field starts at $\Upsilon = \gamma B/B_0 >1$
(here $B_0 = \alpha e/r_e^2$ is the critical field).  In the case of
the electromagnetic wave $\Upsilon \sim 0.5 \xi x$. For $x=50$ and
$\xi^2 =3$, we have $\Upsilon \sim 75 \gg 1$.  Besides that, in order
to consider the coherent \EPEM\ creation in the wave using the same
formulas as for the constant field, the formation length for this
process should be shorter than the wave length. This is also fulfilled
for $\xi >1$. If that is so, then one can forget about using large $x$
and $\xi^2>1$ at photon colliders.  But now I am not sure about the
accuracy of the numbers in the transition regime considered, it would
be worth making an accurate study on this subject.

For convenience,  a set of useful approximate formulas for the
conversion region expressed via the f-number( $f_\#$ = focal
length/diameter = $1/2\theta$) is presented below, the
right part of equations is given for $s=2$:
$$
l_{\gamma} \sim  4f_\#^2 s \lambda/\pi +l_e \approx
2.5f_\#^2\lambda +l_e
$$
$$
\sigma_x \approx s\lambda f_\#/2\pi \sim \lambda f_\#/\pi
$$
$$
\xi^2 = \frac{p}{\sqrt{2\pi}\alpha f_\#^2 s}
\left(\frac{\sigma_0}{\sigma_c}\right) \approx \frac{27p}{f_\#^2}
\left(\frac{\sigma_0}{\sigma_c}\right)
$$

\begin{equation}
A \approx 20p \left(\frac{\sigma_0} {\sigma_c}\right) l_e 
\left(1+\frac{5f_\#^2\lambda}{l_e}\right) \;\mbox{J}
\label{A4}
\end{equation}

The last line is equation (\ref{A3}) with corrected coefficients.
 The input parameter for these equations is
$\xi^2$. The third equation gives $f_\#$ which should be substituted
into the other equations.

Obtaining these formulas we assumed that the transverse size of the
electron bunch is much smaller than $\sigma_x$ of the laser beam. This
is not always correct. In the scheme with crab crossing (see
Fig.\ref{ggcol}), the electron beam is tilted by the angle $\alpha_c/2
\sim 0.015$, that is equivalent to the effective transverse size
\begin{equation}
\sigma_{x,e}=\sigma_z \alpha_c /2. 
\end{equation}
This is an additional
constraint for  choice of  $f_\#$. It can be taken into account
in the following way. \\
1) If $\sigma_{x,e}< \sigma_x$ given by Eq.\ref{A4}b, then all remains
the same, just Eq.\ref{A4}d for the energy should be multiplied by a factor of
$\sqrt{1+\sigma_{x,e}^2/\sigma_x^2}$; \\
2) If $\sigma_{x,e} > \sigma_x$, then one should start the calculation from
\begin{equation}
\sigma_{x,e}=\lambda f_\# /\pi,
\end{equation}
this gives $f_\#$, which should be substituted in eqs.\ref{A4}a,c,d.
Besides that,  Eq.\ref{A4}d should be multiplied by a factor of
$\sqrt{2}$.

\section{Conversion region with a ``traveling laser focus'' \label{new}}

In the previous section we have considered the ``usual'' method of laser beam
focusing.  In this case the laser energy is  not effectively used. In
order to get high conversion probability at a fixed value of the
parameter $\xi^2$, the length of the laser target and the diameter of the
laser beam should be large enough, most of laser photons do not cross the
electron beam.  This is the reason why the required flash energy grows
with the increase of the laser wave length.

\begin{figure}[!hbt]
\centering
\vspace*{0.1cm}
\epsfig{file=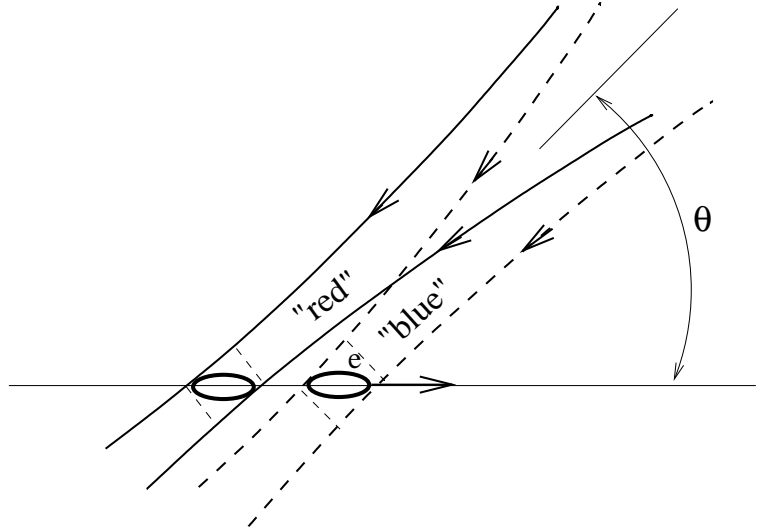,width=10cm,angle=0}
\vspace*{-0.cm}
\caption{Traveling laser focus at the conversion region.}
\label{travfoc}
\end{figure}

Fortunately, there is a way to overcome this problem, that is {\it
  traveling laser focus}~\cite{TSB1}, see Fig.\ref{travfoc}.  In this
scheme, the laser beam follows the electron beam. This can be done using
chirped laser pulses (the wave length changes linearly along the
bunch). Chirped pulse technique is used for stretching and compression of
laser pulses in practically all powerful  short-pulse  lasers.  In order
to make a traveling focus the beam should be pre-focused using elements
with some chromaticity, (Fresnel lenses, dispersive lenses, wedges,
gratings). The picosecond pulses which are needed for photon colliders
have naturally quite broad spectrum and it is not a big problem to vary
the focal length on about $\Delta f/f \sim$ mm/m $\sim 10^{-3}$ and change the
direction on $\Delta \theta \sim d \Delta f/f^2 \sim 10^{-5}-10^{-4}$.
The optimum diameter of the laser beam should be approximately equal
to $2\sigma_x \sim 2\sigma_z \theta$.  The length of the laser beam
can now be arbitrary, it should only be shorter than the distance
between the conversion and interaction points which is about $\gamma
\sigma_y$ or 3.7 mm for $2E=5$ TeV with $\sigma_y = 0.75$ nm (CLIC). In
fact, now the effective thickness of the laser target is equal to the laser 
bunch length. 
For obtaining a high conversion probability in such a scheme, one can use a 
laser with much smaller flash energy and have much smaller values of
the parameter $\xi^2$.

Let us calculate $A$ and $\xi^2$. The laser bunch radius
\begin{equation}
\sigma_x \sim \sigma_z\theta.
\label{sx1}
\end{equation}
The probability  of scattering for the electron is
\begin{equation}
p \sim (2/\sqrt{2})n_0\sigma_c l_{\gamma},
\label{sprob}
\end{equation}
here factor the 2 is due to the relative motion and $1/\sqrt{2}$ due to the
collision of the Gaussian beams with equal radius.
Substituting (\ref{theta}) into (\ref{sx1})  we get
\begin{equation}
\sigma_x = \sqrt{\frac{\lambda \sigma_z s}{4\pi}}.
\label{sx2}
\end{equation}
Substituting (\ref{sx2}),(\ref{sprob}) into (\ref{A1}) we obtain the 
flash energy
\begin{equation}
A=\frac{\pi\hbar c \sigma_z sp}{\sqrt{2}\sigma_c}.
\label{sA1}
\end{equation}
Assuming $s=2$, $p=1$, $l_e \approx 2\sigma_z$ we obtain
\begin{equation}
A \sim 2.2 \frac{\hbar c l_e}{\sigma_c} \sim 
28 l_e[\mbox{cm}]\left(\frac{\sigma_0}{\sigma_c}\right)\;[\mbox{J}].
\label{sA2}
\end{equation}
The value of $\xi^2$ follows from (\ref{xi^2}) and (\ref{sprob}) 
\begin{equation}
\xi^2 =\frac{\sqrt{2}\lambda}{\pi \alpha l_{\gamma}} 
\left(\frac{\sigma_0}{\sigma_c}\right)p.
\label{sxi2}
\end{equation}

One remark. All this is valid for infinitely thin electron beams. In
the case of the crab crossing the effective radius of the electron
beam is $\sigma_{x,e}=0.5\alpha_c \sigma_z$. The above formulas in this
section are valid when $\sigma_x$, given by Eq.\ref{sx2}, is larger than
$\sigma_{x,e}$. For $s=2$ this gives
\begin{equation}
\sigma_z < \lambda/\alpha_c^2.
\end{equation}
For $\alpha_c = 0.03$ this gives $\sigma_z < 1000 \lambda$, that is valid
practically always.

Let us consider, for example, the case b) from table \ref{tflash}.
Taking $l_{\gamma} = 0.3$ cm (we discussed this in the second
paragraph of this section), $\lambda = 10$ \MKM, $\sigma_c/\sigma_0 =
0.7$, $l_e = 0.02$ cm we obtain
\begin{equation}
A \sim 0.8\;\mbox{J}, \;\;\;\;\;\;\;\;\xi^2 \sim 0.3.
\label{example}
\end{equation}
This should be compared with $A=8.3$ J and $\xi^2=1$ for the usual focusing
(see table \ref{tflash}).  To obtain   $\xi^2 =0.3$, using the usual
 method  of focusing, one would need 25 J flash energy.

So, the traveling focus is a very attractive solution for one of the most
serious problems of multi-TeV photon colliders.  Certainly, this
method of focusing is more complicated, but it can significantly reduce
the required flash energy and solve the problem of nonlinear effects.
The chirped pulses needed for this method can be obtained, not only
with usual lasers, but also with free electron lasers.

\section{Interaction region aspects}

\subsection{Collision effects}

Photons are neutral particles, nevertheless there is one collision
effect which restricts the \GG\ luminosity, that is the coherent pair
creation --- conversion of high energy photons into \EPEM\ pairs in
the field of the opposing electron beam~\cite{CHTEL,TEL90,TEL95}.
Detailed study of this limitation was given in Ref.\cite{TSB2}.  At 
high energies, one cannot just collide electron beams and convert
them to high energy photons, as is possible for low energy photon colliders
where \LGG\ is determined only by the geometric electron-electron
luminosity.

Let me compare, just for illustration, the probabilities of
beamstrahlung and coherent pair creation in a strong electromagnetic
field. For $\Upsilon \gtrsim 20$ (in \EPEM\ collisions in the CLIC (5
TeV) project $\Upsilon \sim 20$) the probabilities of these processes
per unit length are $p_{e\to e\gamma} =
5\alpha^2\Upsilon^{2/3}/(2\sqrt{3}r_e\gamma)$ \cite{YOKOYACH} and
$p_{\gamma \to e^+e^-} =
0.38\alpha^2\Upsilon^{2/3}/(r_e\gamma)$\cite{CHTEL,TEL90}. The ratio
of beamstrahlung/pair creation is about 3.8. At \EPEM\ colliders the
average number of emitted photons is usually 1--3, so at photon
colliders the conversion probability to \EPEM\ pairs will be the same
 at somewhat smaller horizontal beam size.

\begin{figure}[!hb]
\centering
\vspace{4mm}
\epsfig{file=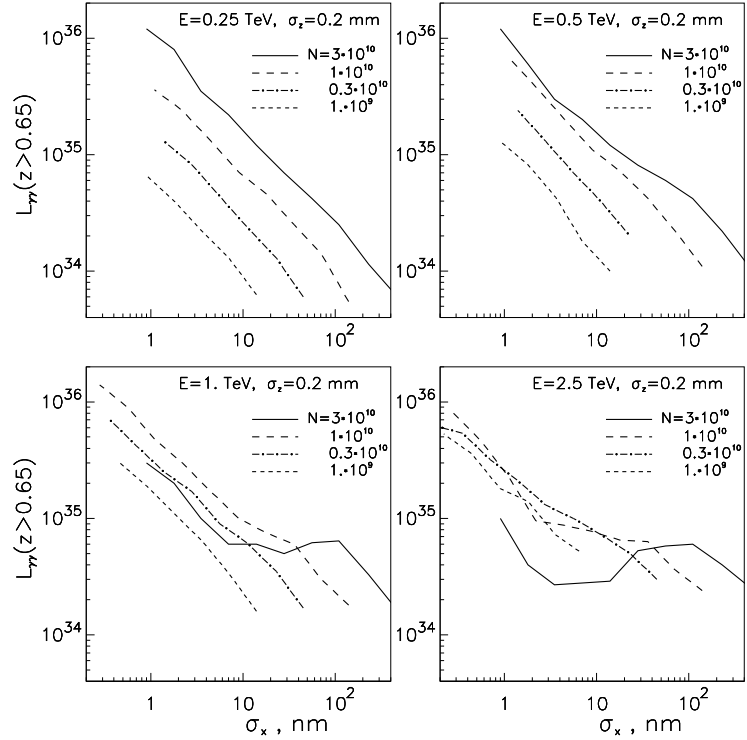,width=10cm,bb= 32 50 550 540 }
\vspace{2mm}
\caption{Dependence of the \GG\ luminosity on the horizontal beam size
for $\sigma_z = 0.2$ \MM, see comments in the text.}
\vspace{-2mm}
\label{lumi}
\end{figure}

In Fig.~\ref{lumi} from Ref.\cite{TSB2} the dependence of the \GG\
luminosity in the high energy peak on the horizontal beam size at
various energies and numbers of particles in the electron bunches is
shown. It was assumed that $N \times f =10^{14}$, the electron
vertical beam size is somewhat smaller than the photon beam size due
to Compton scattering (equal to $b/\gamma$) and the distance, $b$,
between the conversion and interaction points was taken as small as
possible (but sufficient for conversion) $b=3\sigma_x +0.04E_0[\TEV]$,
cm.
In this figures one can see that some curves follow $\LGG \propto
1/\sigma_x$  as is expected in the absence of collision effects, while
some curves make zigzags due to the coherent pair creation. 

For CLIC ($2E_0=5$ TeV) with $N \times f =0.3\times 10^{14}$ the maximum
\GG\ luminosity at a reasonable $\sigma_x$ is about $2\times10^{34}$ \CMS.
Note, this result very weakly depends on $\sigma_z$.
So, the luminosity is not too large, only about a factor of two larger than
that at the TESLA collider at $2E=500$ GeV~\cite{TELTESLA}.

At low energies the coherent pair creation is considerably reduced
(even suppressed) due to the beam repulsion~\cite{TSB2}, see
Fig.\ref{lumi}a. At  high energies this effects works only for long
electron bunches and a small number of particles in the bunch. In
Fig.\ref{lumi} for 2E = 5 TeV this regime corresponds to the region of
very small $\sigma_x$, where the luminosity is really large. This regime, with
small numbers of particles, high collision rate and very small
$\sigma_x$, is practically impossible to reach in reality.

\subsection{A  way to avoid coherent pair creation}

Below I would like to present a new idea on how the problem of the coherent
pair creation can be overcome in photon colliders (at least in
principle). 

The field of the electron beam at the interaction point is
\begin{equation}
B_b = |B|+|E| \sim \frac{eN}{\sigma_x\sigma_z}.
\label{B}
\end{equation}
The coherent pair creation occurs when $\Upsilon =\gamma B/B_0 > 1$~
($B_0=\alpha e /r_e^2)$ and the conversion probability  grows
with the increase of $\Upsilon$ \cite{CHTEL,TEL90}.  For low energy
colliders the field at the interaction point in ee collisions is lower
than $B_b$ given by (\ref{B}) due to the beam repulsion. As result, at
the energies below about $2E_0=500-800$ GeV the coherent pair creation
is not essential even for very tight electron
beams~\cite{TSB2,Tfrei,TELTESLA}. At multi-TeV energies the deflection
is not sufficient for suppression of coherent pair creation.  What can
be done (besides, the old idea of neutralization of the beam field
using four beam \EPEM\ -- \EPEM\ collisions) about this?

\begin{figure}[!hbt]
\centering
\vspace*{0.2cm}
\epsfig{file=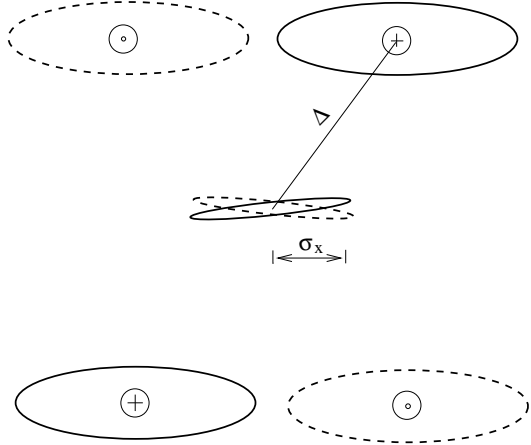,width=7cm,angle=0}
\vspace*{0.2cm}
\caption{Idea of suppression of coherent pair creation at photon 
colliders.}
\label{suppression}
\end{figure}

One idea is the following. Let us collide electron beams tilted
around the collision axis by some relative angle $\phi \sim {\mathcal O(0.1)}
\sigma_y/\sigma_x$. Having initial displacement each of the electron beams
will be split during the collision in two parts, see Fig.\ref{suppression}.
If the transverse deflection during the beam collision, $\Delta > \sigma_x$,
then the maximum field in the region of high energy photons
(at $x \sim \sigma_x$) is
\begin{equation}
B_r = B_b (\sigma_x/\Delta)^2 \propto \sigma_x.
\label{Bs}
\end{equation}

So, with the decrease of $\sigma_x$, the \GG\ luminosity grows but
the field at the interaction region is decreased! 

The main problem here is the production of beams with sufficiently small
sizes.  The required horizontal beam size is larger for longer beams
(the dependence is between linear and quadratic).  For the bunch
lengths, $\sigma_z \sim 25$ \MKM, now considered  for CLIC, this idea,
certainly, does not work. But why should $\sigma_z$  be so short for
photon colliders?  Short bunches are needed during the acceleration to
reduce wake fields,~ \footnote{and to reduce instabilities in collisions of 
\EPEM\ beams which is not necessary in the case of photon colliders}
but after acceleration one can stretch the beam
(using energy spread).  

So, the idea is interesting and worth more detailed consideration. 
 Further study and optimization should
be based on full simulation.

\section{Conclusion, acknowledgments}

Multi-TeV photon colliders have many specific problems, several of them were
considered in this paper, but the main study and R\&D are still ahead.   

I appreciate efforts of the CLIC team towards the multi-TeV linear
collider and their desire to have a second interaction region for \GG\ 
and \GE\ collisions.


\end{document}